# SafeMPI - Extending MPI for Byzantine Error Detection on Parallel Clusters

Dmitry Mogilevsky, Sean Keller

4th July 2018

**Abstract**

Modern high-performance computing relies heavily on the use of commodity processors arranged together in clusters. These clusters consist of individual nodes (typically off-the-shelf single or dual processor machines) connected together with a high speed interconnect. Using cluster computation has many benefits, but also carries the liability of being failure prone due to the sheer number of components involved. Many effective solutions have been proposed to aid failure recovery in clusters, their one significant downside being the failure models they support. Most of the work in the area has focused on detecting and correcting fail-stop errors. We propose a system that will also detect more general error models, such as Byzantine errors, thus allowing existing failure recovery methods to handle them correctly.

## 1  Introduction

In recent years, building clusters of cheap, commodity processors for high-performance computation has become the dominant alternative to using expensive, high-performance processors. At present, clusters dominate the supercomputing market - the latest TOP500 list confirms that the fastest 10 computers in the world are all built using the cluster paradigm, the highest performing at a maximum speed of over 90 TFlops[1]. Clusters are now used extensively for scientific computations by organizations such as Lawrence Livermore National Laboratory, National Center for Supercomputing Applications and the Los Alamos National Laboratory, as well as for business applications by companies such as Google[2].

### 1.1  Trade-off for speed - Cluster failure rate

One commonly cited problem with clusters built from commodity hardware is the component failure rate. While single component failure rate may be acceptable for individual use - having a MTTF of thousands of hours, in a machine that employs a very large number of such components, such as IBM BlueGene, one may expect to see a component failing every day[3]. This represents a difficulty due to the nature of cluster computing, where each individual node is responsible for some essential part of the computation. The cluster community is well aware of this problem and many error recovery protocols have been proposed and developed, such as rollback recovery, log-based recovery (pessimistic, optimistic and causal) and replication[4] . Packages such as MPICH-V[5], Charm++[6] and others offer support for fault-tolerant computing environment.

## 1.2 Fault Models

In order for a fault-tolerant environment to be effective, it must support a good mix of fault models. Currently, three fault models have been proposed.

### 1.2.1 Fail-Stop Faults

This model represents the simplest fault to diagnose and is also the one most frequently assumed by existing fault-tolerance packages. Under this model, a node malfunction can easily be detected by the other nodes as it stops producing output. This model corresponds naturally to events such as complete hardware failure (possibly due to power source failure), or complete network connectivity failure.

### 1.2.2 Fail-Stutter Faults

This model is a natural extension of the Fail-Stop model. It assumes that a node may 'stutter', i.e. experience a drop in performance (full or partial) for a period of time, then return to normal operation. This model is significantly more expressive than the Fail-Stop model, as it addresses node availability in terms of performance, as well as functionality[4].

### 1.2.3 Byzantine Faults

This is an extremely adversarial fault model which assumes that a node, or any number of nodes may choose to engage in malicious behavior. Under this model, no assumptions are made about the ability of other nodes to detect a fault. In fact, it is most commonly assumed that a node will continue providing output to the other nodes, but the output will be incorrect. Faulty nodes under this model are allowed to collude with each other in order to act in a more malicious way.

## 1.3 Motivation - Detecting Faults

Currently, most fault-tolerance packages deployed either assume a Fail-Stop failure model or assume a perfect fault-detecting mechanism. In other words, their focus is mostly on correcting errors that have been detected. Yet, it has been suggested that the Fail-Stop model is extremely constricted and does not come anywhere near reflecting the real situation with regards to faults on computational nodes in clusters[7, 4]. Detecting Byzantine faults is significantly more difficult than detecting Fail-Stop faults due to their insidious nature. In order to attempt to detect them, it may be helpful to further divide such errors into two categories -

- Transient faults - Non repeating errors which corrupt the data by reversing some bits. These errors are typically caused by situational events such as incidental electromagnetic interference, 'ground bounce' and external radiation[8].

- Permanent faults - Faults caused by hardware or software defects (due to aging, external damage, or sabotage), which will cause error in computation to repeat across executions.

One approach that has been suggested for detection of Byzantine errors is to use error-detecting versions of commonly computer algorithms such as matrix operations [12] and Fast Fourier Transform data compression [10]. While this is a viable approach, it is more reactive than proactive. In particular, it requires creating fault-tolerant versions of commonly performed computations, incurring significant overhead in time and resources required to perform the computation. We feel a different approach should be taken to detecting permanent fault. Our approach is to diagnose misbehaving nodes in the cluster and report their behavior to the agent in charge of error recovery, so that the necessary steps, such as data roll-back and relocating the computation to a different node can be taken by the system.

## 2 Diagnosing Faults By Challenging the System

Instead of customizing individual algorithms to tolerate faults, we propose to create a parallel application which will detect which parts of the system function incorrectly through an algorithm in which the nodes seek to derive a solution to a fairly light-weight problem representative of the cluster workload. The results of the computations are then compared to each other. Based on that comparison, a decision is made whether or not certain nodes behave in a manner which is not correct. Our algorithm will support detection of faults under all three models presented in section 1.2 of this report. As a side effect, it will also detect any problems existing on the communication interconnect, which previously has been assumed to be fault-free. We see the following design criteria as essential elements for our fault-detection algorithm

- Lightweight - Above all, the use of the application should not significantly degrade the performance of regularly scheduled tasks on the cluster. While this may seem like a tough requirement to achieve, it is not necessarily so, since diagnostics need not be run at very high frequency.

- Non-deterministic - The application should use a randomized data set in order to eliminate the possibility of fake responses

## 2.1 The Byzantine Generals Problem

Detecting Byzantine errors within a large computing cluster is not a simple task, and any proposed algorithmic solution needs to be backed by rigorous mathematical proof. Fortunately, coping with Byzantine faults in a cluster can be reasoned about abstractly as the Byzantine Generals Problem. In this abstraction, every node in a cluster is represented as a soldier in the Byzantine army. Any soldier who makes a decision about what action to take is called a general and all other soldiers are lieutenants. Communication between soldiers is accomplished by passing messages. The soldiers will communicate with each other to form a common plan of action (attack or retreat), but some of the soldiers are loyal and some are traitorous. The traitorous soldiers may be trying their hardest to prevent the loyal generals from reaching a common agreement. An algorithm is needed so that all loyal generals decide upon the same plan of action and a small number of traitors cannot cause the loyal generals to adopt a bad plan[13].

### 2.1.1 The Byzantine General Problem as formulated by Lamport et al[13].

A commanding general must send an order to his n-1 lieutenant generals such that:
    IC1: All loyal lieutenants obey the same order
    IC2: If the commanding general is loyal, then every loyal lieutenant obeys the order he sends.

### 2.1.2 Solution Constraints

Lamport et al.[13] offer an elegant solution to this problem which provides detection of up to m traitors given $3m + 1$ or more generals. The solution uses a message system which has the following constraints:
    A1: Every message that is sent is delivered correctly.
    A2: The receiver of a message knows who sent it.
    A3: The absence of a message can be detected.

### 2.1.3 A solution to the Byzantine General Problem: OM(m)

- m represents the number of soldiers.

- The possible actions are ATTACK or by default RETREAT.

**Algorithm OM(0).**

1. The commander sends his value to every lieutenant.

2. Each lieutenant uses the value he receives from the commander, or uses the value RETREAT if he receives no value.

**Algorithm OM(m), m > 0.**

1. The commander sends his value to every lieutenant.

2. For each i, let vi be the value Lieutenant i receives from the commander, or else be RETREAT if he receives no value. Lieutenant i acts as the commander in Algorithm OM(m - 1) to send the value vi to each of the n - 2 other lieutenants.

3. For each i, and each j ~ i, let vj be the value Lieutenant i received from Lieutenant j in step (2) (using Algorithm OM(m - 1)), or else RETREAT if he received no such value. Lieutenant i uses the value majority (vl . . . . . v,-1 ).

## 2.2 The System Wide Detection Algorithm

First, we reduce the problem of Byzantine fault detection to the Byzantine Generals problem. Then, we provide a system wide distributed algorithmic solution which reduces to the OM(m) algorithm. By showing equivalence in this way, our algorithm is proved to be sound and provide error detection of up to m node failures in a cluster of at least 3m+1 nodes.

Given an arbitrary network of N nodes we let each node maintains a table of reliability estimates for every node in the system. We construct a virtual token-ring network where a single token is passed around the network sequentially and cyclically from node 1 to 2 to ... to N -1 to N to 1. The token is passed periodically at a predetermined frequency; the period of this rotation is referred to as the epoch time. The token has a payload consisting of: a counter which is incremented every time it is passed, a reliability table similar to the one held by every node, the sender's ID, the receiver's ID, and a checksum over the data. The current token holder is analogous to the general and all other nodes the lieutenants. When a node G holds the token it is responsible for performing the network wide challenge. Node G generates a challenge and broadcasts the challenge to all other nodes (the challenge algorithm is detailed in section 2.5). The other nodes respond to node G and node G uses the results of the challenge to update its reliability table. Node G then passes the token to the next node in the ring, node G+1 (where +/- is closed within the ring modulo N), and simultaneously broadcasts its reliability data to every other node in the network. This is accomplished with a a single token broadcast. When the other nodes receive the broadcast, they compare the reliability data generated by node G to their current reliability data. If the change in each value does not correspond to the expected change of each data being deterministically raised or lowered then the table is considered corrupt, and the local reliability table is not updated. This is necessary as we must assume that incorrect data is the result of a traitorous node attempting to corrupt the reliability tables of the loyal nodes.

In order to satisfy the message system constraints put forth by Lamport et al.[13], some problematic situations must be addressed. First, if node G is a traitor, it may not pass the token along to node G+1. To account for this, every node has a timer which is reset when it receives a token. When node G-1 broadcasts a token successfully, all nodes note the success of node G-1 and reset their timer. If node G fails to broadcast the token, node G+1 will timeout one period after the node G-1 broadcast. At the timeout, node G+1 will notice that the last broadcast it received was from node G-1, and node G+1 will perform the challenge followed by the next token broadcast. Node G+1 will also reduce the reliability estimate for node G in its reliability data. All other nodes will notice that their own ID does not equal G-1+2=G+1, so they will reset their timers and expect a challenge and token broadcast from node G+1 next. If node G+1 fails to generate reliability broadcast and passes the token then node G+2 will notices this failure and generates a challenge and token broadcast. This process can occur for any length sequence of nodes in the ring without causing a communication failure or breaking the consistency model. Additionally, if some node G sends out a challenge, but does not broadcast a token, then node G+1 will timeout one period after G-1 performed the token broadcast and proceed as detailed above.

Another problem that may occur is when a traitorous node generates a spurious broadcast out of order. This could happen if node G becomes very slow, thus allowing the next node in the ring to time out and generate a new token broadcast, but node G finishes its work and generates a broadcast after node G+1. When this occurs, every node will ignore the incorrect broadcast. The nodes will know to ignore the broadcast because the expected sender ID does not match the received sender ID.

Finally, a traitorous node may fail to respond to a challenge. This is countered by the current head, node G, timing out after a predetermined length of time following its challenge broadcast. After broadcasting a challenge, node G will either receive all of the challenge responses or timeout waiting for traitorous nodes to respond. Node G then updates the reliability table penalizing those nodes which failed to respond and broadcasts the next token. Assuming that messages can be guaranteed with a checksum, section 2.4 details how the algorithm satisfies the requirements of a messaging system as defined by Lamport et al.[13].

## 2.3 Algorithmic Equivalence to OM(m)

### 2.3.1 The commander sends his value to every lieutenant.

This is achieved via the token broadcast. The current token holder broadcasts his updated reliability table to every other node, as detailed in section 2.2

**2.3.2** For each i, let vi be the value Lieutenant i receives from the commander, or else be RETREAT if he receives no value. Lieutenant i acts as the commander in Algorithm OM(m - 1) to send the value vi to each of the n - 2 other lieutenants.

Every node that receives the token broadcast updates its local reliability table with the table embedded in the token. If a node does not receive the broadcast, it will time out and leave its local reliability table unchanged. This is equivalent to the default action RETREAT. Upon a token broadcast, the node sequentially following the current token holder in the ring becomes the head node and repeats the algorithm.

**2.3.3** For each i, and each j ~ i, let vj be the value Lieutenant i received from Lieutenant j in step (2) (using Algorithm OM(m - 1)), or else RETREAT if he received no such value. Lieutenant i uses the value majority (vl . . . . . v,-1 ).

After the head node sends out a challenge broadcast it waits for responses until all are received or a time out occurs. The head node then tallies the responses and chooses the majority response as the correct response and updates its local reliability table based on this value. This is a directly analogous to 2.3.3.

## 2.4 Conforming to Lamport's Message System Constraints

### 2.4.1 A1: Every message that is sent is delivered correctly

The checksum over every message probabilistically ensures that in the unlikely event of message data corruption, any recipient will ignore the message. The reliability of the sender will be reduced by any recipients of the corrupt message, so the malformed message is observed by recipients as if sent by a traitorous node. If message corruption persists due to permanent network failure, the recipients will eventually reduce the reliability of the sending nodes below the reliability threshold and the affected nodes will be marked as traitorous. This does not precisely satisfy constraint A1, but it is sufficient in capturing the correct notion of reliability in a physically realizable system. When performing cluster based computation, it is essential to alert the user to any failures, network based or node based. Our algorithm simply aggregates both node and network failure types into general failures. In the presence of transient network failure, the reliability of a node will only be marked down during the transient failure. Its reliability will rebound as soon as the transient fault disappears. Similarly, permanent link failure is observably no different then a faulty node failing to respond. The solution to this problem is detailed in section 2.2. Therefore, assuming that a checksum is sufficient to detect network based data corruption, and network corruption and link failure are observably no different then the same behavior resulting from faulty nodes, every message that is sent is delivered correctly, or is observed as node failure. The consequence of this is that failing network links resulting in K observed node failures reduce the detection of at most m failing nodes in a network of size 3m+1 to the detection of at most |m-K| nodes in a 3m+1 node network.

### 2.4.2 A2: The receiver of a message knows who sent it.

Every message that is sent includes the ID of both the sender and the receiver, so a receiver always knows the reported ID of the sender from the received message. The only uncertainty arises when a sender lies about its identity; a sender can simply write the ID of another node into the sender field of the message. This problem is easily remedied by employing a basic public key cryptography system. First, every node must be provided with the public key of every other node in the system. Second, a sender must encrypt a portion of every message before transmission, and every message recipients must use the public key corresponding to the claimed sender's ID to attempt decryption of the data. Finally, if the message contents can be decrypted then the sender's authenticity is guaranteed to some very high probability determined by the strength of the cryptographic system. We have not implemented this system, but doing so is very straight forward. Instead, our working model does not allow for nodes to lie about their identity which ensures that the receiver of a message knows who sent it.

### 2.4.3 A3: The absence of a message can be detected.

This is ensured via the timeout scheme detailed in section 2.2. In summary, when a message is expected but not received before a predefined duration of time, a timeout occurs and the algorithm makes forward progress ignoring the lost message.

## 2.5 The Challenge Algorithm

The challenge response system must provide a means of discovering failing nodes and marking them as faulty. At the simplest, every node should have a broad range of computation to perform. The computation should be broad enough to include various register and memory accesses, every possible operation within the FPU, and a broad range of operations performed by the ALU. Each operation should depend on the result of the previous operation, and the final result should be a scalar value. This helps to maximize the probability of discovering a failing node if only a small portion of the processor is broken, such as an FPU divider.

    The current general, node G, is in charge of generating one pseudo-random data set for every node to run the challenge on. Upon receiving a token pass and becoming the general, node G waits one epoch time before transmitting the challenge broadcast. The other nodes receive the broadcast, perform the challenge algorithm, and then send the computed result back to node G. Any node which does not reply before a timeout period is considered to be faulting. Node G compares all returned results, including its own results of the challenge

and selects the majority value to be the correct value. Node G then updates its reliability table reducing the reliability of all nodes which failed to respond and all nodes which did not return a value equal to the majority value. Node G also increases the reliability of all nodes which agree with the majority value.

We have chosen to implement the challenge algorithm as a series of matrix operations. Specifically, both the floating point and the integer matrix undergo gaussean elimination into reduced row form. Once in reduced row form, the rows of the matrix are summed together, while the columns are multiplied[1]. The resulting column products are then summed together, as are the resulting row sums. For each matrix, the quotient of the sum of the products and the sum of the sums is derived. Finally, the quotient of the floating point matrix is divided by the quotient of the integer matrix to arrive at a single number. This sequence of operations was selected because it can be performed relatively fast, is widely inclusive in the operations required to perform it, and given different data, converges to different results. Additionally, its use of matrix reduction is typical of a scientific computation workload on a cluster, which frequently deals with matrix operations.

The ideal algorithm for performing the reliability update is both system and application dependent. The simplest method is to pick a large constant d and let the reliability range from completely reliable 1.0 to completely unreliable 0.0. When a node's reliability should be decreased it is decreased by $1/d$, and when it should be increased it is increased by $1/d$. The value is saturated at 0.0 and 1.0 so any increases above 1.0 or decreases below 0.0 are ignored. More advanced solutions allow for weights varying by the current reliability. Finally, when the reliability of a node falls bellow a threshold, the node is considered faulty.

## 2.6 Detecting Fail-Stutter and Fail-Stop Faults.

It should be evident that the sets of all possible fail-stutter and fail-stop faults are subsets of the set of all Byzantine faults, so detection of all Byzantine faults by our algorithm includes detection of all fail-stutter and fail-stop faults.

## 3 The API

Fault tolerance can be added to a system of computers at the application programming level, at the programming language level, at the operating system level, or at the network hardware level. Adding Byzantine error detection at a network level would require a great deal of high speed custom hardware, making it an impractical solution. However, it would be possible to use our algorithm to detect faults by writing support into an operating system. The problem with this approach is that it severely limits the usefulness of the application. We would need to chose a single OS to integrate support into, and then users would need to run our OS to use the software. Adding support into a programming language also severely limits the usefulness of our application, because only our compiler could be used to generate the correct code. The best way to provide support for a wide range of target architectures, operating systems, and networks is to provide the programmer with a useful API to interface with out application.

---

[1]Ignoring zeros, for the obvious reason.

One of the most universally accepted and used parallel computing paradigms is MPI for C. There are many publicly available open source MPI packages, and they are easy to use, modify, and add support to because they all conform to the same standard interface, MPI. We have extended parts of MPI in order to add support for our fault detection algorithm. In this way, programmers will be able to easily add complex fault detection support to their applications by using our API, SafeMPI, an extension of MPI.

# 4 Description of Implementation

## 4.1 Transparency

One of the stated goals of developing a API-based fault-detection algorithm is providing functionality that is transparent to the user. Using this requirement as a guideline, we implemented only three public function calls which provide the user with access to the functionality necessary to make SafeMPI a useful tool. These functions are the initialization function `MPI_Safe_init`, the termination function `MPI_Safe_Finalize` and the reliability score reporting function `Safe_get_score`. `MPI_Safe_Init` and `MPI_Safe_Finalize` need to be called only once by the user program in place of standard `MPI_Init` and `MPI_Finalize` calls. `Safe_get_score` must get called every time the user wishes to perform reliability checks or comparisons in their code. Additionally, the user must declare an instance of data type `safe_struct` which is passed to every SafeMPI call.

## 4.2 Data types

All communications performed by SafeMPI will be done using typedef message types, which are mirrored with defined MPI data types. The encapsulating data type used for all messages is `message_type`, defined as,

```
struct message_type {
  header_type head;
  body_type body;
};
```

The type `header_type` contains the header information of the message, with the following structure,

```
struct header_type {
  unsigned int message_class;
  unsigned int myid;
```

```
  unsigned  int checksum;
};
```

`message_class` indicates the type of the message being communicated (Token, Challenge, Response). `myid` contains the group id of the sending process. `checksum` is used for future implementation of message checksums. The type `body_type` is in fact a union containing the contents of the three possible message types,

```
union body_type {
  token_type tok;
  challenge_type c_data;
  response_type r_data;
};
```

Each of these is defined as,

```
struct token_type {
  unsigned int token_id;
  int cur_token_holder;
  int new_token_holder;
  r_table_type r_table;
};

struct challenge_type {
  float random_float_set[10][10];
  int random_int_set[10][10];
};

struct response_type {
  int source;
  float response;
};
```

The data type `r_table_type` is another structure containing the reliability table in array format.

## 4.3  Initialization and Background Operation

Once user issues a call to MPI_Safe_Init, MPI_Init_thread is called to initialize MPI operation. A child thread is then created to run background reliability checks in, while the main thread starts the execution of the user program. All SafeMPI functionality is executed in this separate thread and using an MPI communicator other than `MPI_COMM_WORLD` to avoid interference with user application. Once initialized, SafeMPI will continue executing in a loop until halted by the user application. In each iteration of the loop, the following operations will be performed on the current round's head node:

1. Generate a 10x10 float and integer random matrix.

2. Create an instance of `challenge_type` with the generated data.

3. Call MPI_Bcast to broadcast the data to the nodes in the job group.

4. Wait until all responses have been received and stored in the responses array, or until a response timeout occurs.

5. If a response timeout occurs, set each node that timed out to the default wrong value.

6. Sort responses and select the median response as correct.

7. For each node, if its response matches the selected response, compute the new reliability score using the formula $new\_score = min(1, old\_score + \delta^{\frac{1}{1-old\_score/1.5}})$, where $\delta = 0.2$. If the response does not match the selected response, compute the new reliability score for that node using the formula $new\_score = max(0, old\_score - \delta^{\frac{1}{old\_score1.5}})$, where $\delta = 0.2^2$.

8. Generate a new token message containing the new reliability table and `new_token_holder` with the value `(myid + 1)%size`, where `myid` is the process id of the current head node and `size` is the size of the job group.

9. Wait for the period of time corresponding to Epoch Time.

10. Call MPI_Bcast to broadcast the token to every node in the job group.

All nodes that are not the head node in a particular round will perform the following operations:

1. Spawn a child thread `child_thread` to listen to broadcasts and forward them to the parent thread.

2. Poll the communicator pipe with `child_thread` for forwarded messages for a specified period of time.

3. When data is detected on the pipe, store the message and check its `message_type` field.

4. If the message type is a challenge, then store the challenge data and use it to compute a response value using the `generate_response` function. Once a response is computed, create a `response_type` message and use MPI_Send to send it to the head node.

---

[2]These particular functions were selected due to their exponential nature. This ensures that a node which has a high reliability score and starts producing bad results will have its score lowered very quickly. Likewise, a node with low score which operates correctly will recover its score to an acceptable level within only a few iterations

5. If the message type is a token, store the received reliability table. For each entry in the table, if $old_value > new_value$, compare $new_value$ with $max(0, old\_score - \delta^{\frac{1}{old\_score1.5}})$. If $old_value < new_value$, compare $new_value$ with $min(1, old\_score + \delta^{\frac{1}{1-old\_score/1.5}})$. If there is a mismatch in the values, discard the table as erroneous.

6. Set the current head node to the value of `new_token_holder`, and leave loop iteration.

7. If a timeout occurs, increment the current head node by one and jump to next loop iteration.

## 4.4 Finalization

Upon calling `MPI_Safe_Finalize`, all open child processes is terminated, and a call to `MPI_Finalize` is issued.

## 4.5 User Reliability Checks

During every iteration of SafeMPI algorithm, after receiving the updated version of the reliability table and verifying it as valid, each node will write the table to a communicator pipe to the parent (user) thread. When the user program issues a call to `Safe_get_score`, the pipe is polled and the data in it is stored in the parent thread. The score for the requested node is then returned to the user.

# 5 Development and Testing

Our development and testing took place on NCSA's Uranium cluster. The Uranium cluster is a 32-node, IBM eServer, Pentium III test cluster administered and maintained by the Security Research division of NCSA. The cluster consists of one head and 31 compute dual processor nodes. Both 100MB Ethernet and Myrinet are used to connect the nodes. After initial analysis and testing, we thoroughly tested the SafeMPI implementation to determine its performance overhead and its reliability data convergence time in a faulty network.

## 5.1 Performance overhead

SafeMPI offers numerous user controllable settings such as the epoch period, and timeout durations. The most important of these settings is the epoch period which sets the frequency of challenge broadcasts. In order to determine how SafeMPI scales with respect to epoch period and the number of nodes within a ring, we ran benchmarks on a practical real-world algorithm, Sobel edge detection. The Sobel algorithm is widely used to detect and enhance the edges found within a digital image. Typically, the image is represented by a matrix of RGB values, and the Sobel operator performs a two dimensional spatial gradient measurement to find and emphasize regions of high spatial frequency which are very likely to be edges[14]. Sobel performs computation on convolution kernels which are simply small subsections of the full matrix representing an image, so the algorithm parallelizes very naturally. The parallel version moderately relies on inter-node data communication and can execute over a predefined number of slowly converging iterations. These properties make the parallel Sobel algorithm an ideal test candidate for use with SafeMPI. The results of the tests are detailed in Table 1.

| Nodes | 20s Epoch Time | 30s Epoch Time |
|---|---|---|
| 6 | 6.04% | 11.43% |
| 12 | 9.47% | 18.63% |

Table 1: Execution time overhead for the Sobel Algorithm using Safe MPI vs MPI.

## 5.2 Reliability report convergence in the presence of failing nodes.

Failing nodes can be simulated within SafeMPI by probabilistically either corrupting a challenge reply message or by failing to send out either: a token broadcast, a challenge broadcast, or a challenge reply. Using the reliability table update mechanism detailed in section 4.3 and a reliability threshold of 0.30, failing nodes are reliably detected within 5 or 6 epochs. This is entirely independent of the use application or other SafeMPI parameters.

# 6 Discussion

Our work on this project has convinced us that the idea of developing a challenge-based fault detection API has significant merit. If nothing else, reliability score reports can be logged during the duration of the MPI job and examined later by the user to alert them of an existing hardware problem, as well as the potential compromise of the results. There are more far-reaching possibilities such as integrating reliability checking with cluster management tools to allow the scheduler to intelligently migrate a job off a failing node to a healthy one if a problem is detected. These measures could be potentially very time saving, as they allow to diagnose a problem provocatively, before a total hardware failure occurs. This is especially desirable in systems which do not implement check-pointing on non-volatile storage.

However, in spite of the promise that the results of this project have shown, we were unable to accomplish everything we initially set out to achieve. Although it is possible to use SafeMPI in a real-world application, as our experiments show, in the current state, the library still contains a number of bugs, the presence of which make its use in real-world situations unrecommended, and which we intend to continue to hunt down and eliminate. During the implementation of our algorithm we have encountered a number of challenges we did not initially anticipate, related directly to the existing cluster technology and MPI standards. Current MPI implementation offer poor support for heavily asynchronous multi-threaded applications that invoke simultaneous MPI calls. That barrier has complicated our work significantly.

Furthermore, in its current implementation, as demonstrated by our results, SafeMPI has serious scaling issues. In its present state, scaling presents too serious of an issue to consider using it in production environment. In order to make SafeMPI a viable tool, we intend to identify as many inefficiencies in the algorithm as possible and eliminate them. One particularly disappointing result is that raising the time between epochs of the algorithm seems to increase execution time as oppose to decreasing it, as one would expect. We suspect that this is not an inherent property of the algorithm, but rather a fault of the implementation. Identifying and eliminating the cause of this occurrence is a very important task on the path toward making SafeMPI a useful tool. Other techniques for improving scalability and performance issues depend more on the modification and optimization of the core algorithm itself.

On the other hand, one particular advantage in our method is that, unlike fault tolerance through thread and hardware level redundancy, our overhead is purely temporal and thus is subject to optimization. Replication, while quite efficient in detection of transient errors, their overhead is bound by the need to fully replicate the user process. This is a hard bound which does not exist in our implementation of fault detection, as better and better optimization of the implementation and algorithm will produce less and less overhead.

# 7 Concluding Remarks

Many solutions have been proposed to aid failure recovery in clusters, but they only support the simplest of failure models. Most of the work in the area has focused on detecting and correcting fail-stop errors, but these errors do not represent the types of problems encountered within a real cluster. In this paper, we have presented a solution to detect errors using the most robust failure model available, the Byzantine fault-model.

Our results show that it is possible to detect errors within this substantially more complex model and the overhead in doing so remains low for small groups of nodes. While our results indicate difficulty in scaling with the number of nodes, future optimizations and minor changes to the global algorithm will most likely yield lower overhead and increased scalability. In the meantime, one promising approach to scaling may be to break the global ring into smaller sub rings; this reduces the error detection rate but makes the current implementation usable in a large cluster.

# 8 Acknowledgments

We would like to thank the Security Group of National Center for Supercomputing Applications for providing us with the resources and facilitates to realize this project.